# X-ray nanoimaging of crystal defects in single grains of solid-state electrolyte Al$_x$Li$_{7-3x}$La$_3$Zr$_2$O$_{12}$

Yifei Sun[1], Oleg Gorobstov[1], Linqin Mu[2], Daniel Weinstock[1], Ryan Bouck[1], Wonsuk Cha[3], Nikolaos Bouklas[4], Feng Lin[2,*], Andrej Singer[1,*]

[1]Department of Materials Science and Engineering, Cornell University, Ithaca, New York 14850, United States

[2]Department of Chemistry, Virginia Tech, Blacksburg, Virginia 24061, United States

[3]Advanced Photon Source, Argonne National Laboratory, Argonne, Illinois 60439, United States

[4]Sibley School of Mechanical and Aerospace Engineering, Cornell University, Ithaca, New York 14850, United States

[*]Corresponding authors, fenglin@vt.edu, asinger@cornell.edu

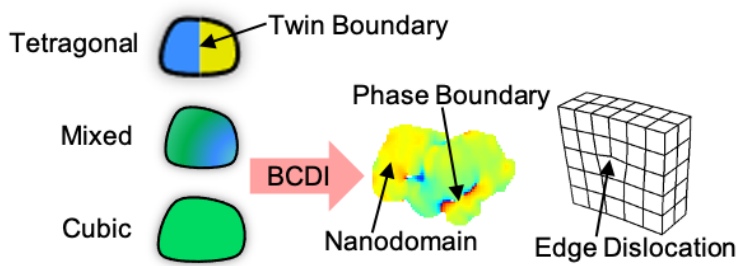

**All-solid-state lithium batteries promise significant improvements in energy density and safety over traditional liquid electrolyte batteries. The Al-doped Al$_x$Li$_{7-3x}$La$_3$Zr$_2$O$_{12}$ (LLZO) solid-state electrolyte shows excellent potential given its high ionic conductivity and good thermal, chemical, and electrochemical stability. Nevertheless, further improvements on LLZO's electrochemical and mechanical properties call for an incisive understanding of its local microstructure. Here, we employ Bragg Coherent Diffractive Imaging to investigate the atomic displacements inside single grains of LLZO with various Al-doping concentrations, resulting in cubic, tetragonal, and cubic-tetragonal mixed structures. We observe coexisting domains of different crystallographic orientations in the tetragonal structure. We further show that Al doping leads to crystal defects such as dislocations and phase boundary in the mixed- and cubic-phase grain. This study addresses the effect of Al-doping on the nanoscale structure within individual grains of LLZO, which is informative for the future development of solid-state batteries.**

The ever-increasing demand for safer, higher density, and temperature insensitive energy storage systems has fueled the development of all-solid-state Li batteries, which allow using the lithium metal anode without the notorious dendrite formation.[1–4] At present, most efforts focus on designing solid-state electrolytes with high ionic conductivity ($> 10^{-4}$ S cm$^{-1}$). Additional considerations include chemical, electrochemical, and structural stability against lithium metal and various cathode materials.[5] There are several potential options for solid inorganic electrolytes, including garnet $Li_7La_3Zr_2O_{12}$ (LLZO), $LiZr_2(PO_4)_3$ (NASCION), LiPON, perovskite $Li_{0.5}La_{0.5}TiO_3$, antiperovskite $Li_3OCl$, and sulfide $Li_{10}GeP_2S_{12}$. Recent research has shown that specific crystal structures, such as the body-centered cubic structure, can allow facile ionic conduction,[1] presumably due to the lithium's direct hops between adjacent tetrahedral sites with low activation energy.[6] Amongst these structures, the garnet-type electrolytes received significant attention because of their accessible synthesis, high-temperature stability, and high ionic conductivity.[3,4,7–9]

The garnet-type LLZO has shown exceptionally high ionic conductivity ($10^{-3}$ to $10^{-4}$ S / cm) and good chemical stability against Li metal making it is suitable for all-solid-state battery applications.[8,11] While the cubic LLZO (c-LLZO) is reported to have high ionic conductivity, the tetragonal structure (t-LLZO) is stable at room temperature with a conductivity two orders of magnitude lower (both structures are shown in Figure 1a).[9,12,13] One reason for the conductivity difference between the cubic and tetragonal structures is the sparsity of lithium dynamical excitations in t-LLZO due to its strong Li ordering, which limits the ionic pathway in the tetragonal structure.[12,14,15] Recent experiments have demonstrated that aluminum doping ($Al_xLi_{7-3x}La_3Zr_2O_{12}$) can stabilize the metastable pure c-LLZO and increase its ionic conductivity by introducing lithium vacancies from the substitution of $Al^{3+}$-ion for three $Li^+$-ions.[7,16]

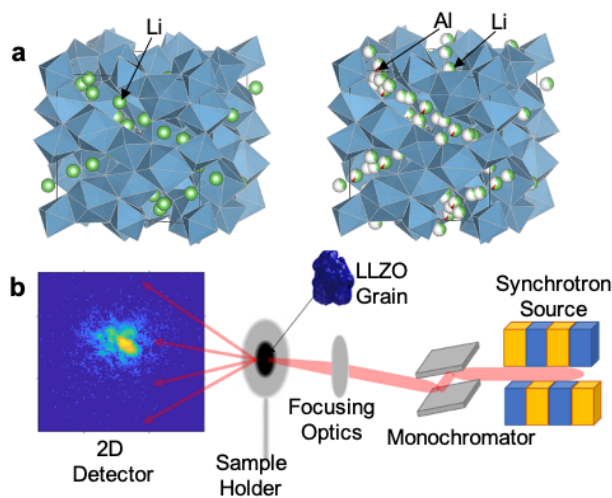

**Figure 1**. (**a**) Coordination polyhedron (blue) around the transition metals La and Zr and the Li sites (green) of the tetragonal (left) and cubic (right) LLZO structure using VESTA.[10] The Al doping (red) is also shown in the cubic structure. The white portion of the Li sites in cubic LLZO indicates a low occupation number compared to the tetragonal LLZO that all Li sites have an occupation number of 1. The occupational disorder in the cubic structure contributes to its high ionic conductivity. The solid black line draws the unit cell. (**b**) The experimental setup and a typical recorded diffraction pattern.

While Al-doping's impact on ionic diffusion received considerable attention, the effect of doping on the microstructure of LLZO warrants a more thorough investigation since it is inherently related to materials' mechanical and electrochemical properties.[17] For example, local strain can dominate the overall ion-transport at the device scale: theoretical studies have predicted that a 5% local strain in LLZO can decrease its conductivity by an order of magnitude.[18,19] The defects such as dislocations and grain boundaries can also contribute to the mechanical failure of LLZO, such as cracking or facilitate dendrite formation.[20] However, the lack of effective tools to probe the microstructure of individual LLZO grains hampered the understanding of the correlation between materials nanoscale structural properties and its ensemble-averaged functionality, which impeded the advancement of more sophisticated solid-state electrolyte design routes.[1,7,8]

Here, we combine single grain x-ray diffraction and Bragg coherent diffractive imaging (BCDI) to study the extended crystalline defects of LLZO grains embedded in sintered pellets (see Figure 1b for experimental setup). Single grain diffraction provides a unique perspective on the microstructure, while BCDI delivers a three-dimensional (3D) displacement field inside a nanocrystal that allows us to study strain, phase distribution, and dislocations within sub-micron grains.[21–23] We report the existence of twin domain boundaries within single LLZO grains deduced from the split diffraction peak of an individual grain in the tetragonal structure. Combined with the 3D imaging of the grain's displacement field, our study shows the development of high-strain regions from insufficient Al-doping in the mixed structure. Imaging also reveals the presence of edge dislocations in both the mixed and cubic structural phases. We expect the observed structural defects to have profound implications on the mechanical and electrical performance of LLZO: while defect sites at the surface are prone to dendrite formation, dislocations in bulk can accelerate the ionic conduction.

The tetragonal - $Al_{x=0}Li_7La_3Zr_2O_{12}$, mixed - $Al_{x=0.13}Li_{6.61}La_3Zr_2O_{12}$, and cubic - $Al_{x=0.24}Li_{6.28}La_3Zr_2O_{12}$ samples are prepared identically through the solid-state synthesis using the precursors of $Li_2CO_3$, $Al_2O_3$, $La(OH)_2$, and $ZrO_2$. The precursors with a certain stoichiometric ratio are thoroughly ball milled at 450 rpm in a $ZrO_2$ jar with isopropyl alcohol (IPA) for eight hours. After being sufficiently dried, the mixed precursors are pressed into pellets using uniaxial pressing in a 10 mm stainless die at the pressure of 10 MPa. The pellets are then transferred into a box furnace, pre-heated at 1000°C for ten hours. At last, the pre-heated precursors are grounded, re-pressed, and re-heated at 1000°C for ten hours.

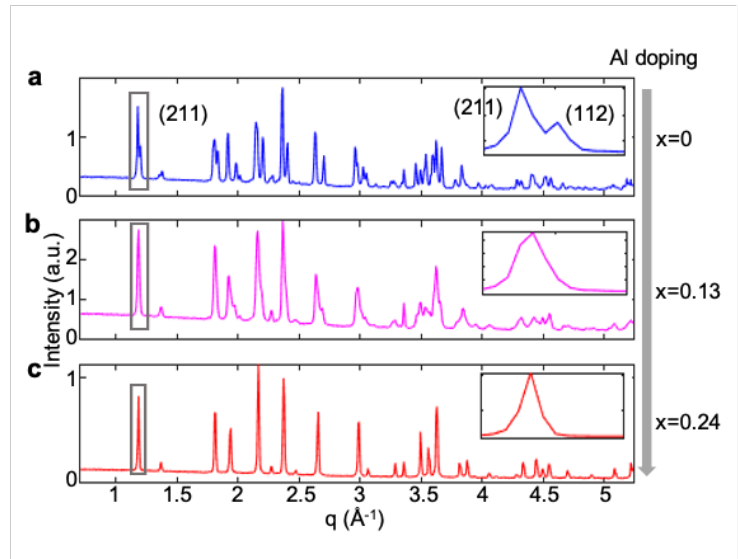

**Figure 2**. Powder XRD patterns for (**a**) tetragonal- $Li_7La_3Zr_2O_{12}$, (**b**) mixed- $Al_{x=0.13}Li_{6.61}La_3Zr_2O_{12}$ (magenta), and (**c**) cubic-$Al_{0.24}Li_{6.28}La_3Zr_2O_{12}$ (red) structures. The insets are the enlarged region of the (211) peak measured for coherent single grain diffraction.

We perform a conventional X-Ray powder diffraction (XRD) measurement to characterize the structures of the three separately synthesized samples (see Figure 2). Consistent with the literature,[16,24] the XRD analysis confirms the presence of t-LLZO at x = 0, characterized by the split diffraction peaks (see inset in Figure 2a). At the largest aluminum content (x = 0.24), the split peak merges into a single peak identified as the cubic structure (see Figure 2c). At the intermediate level of doping, the double peaks merge into broad single peaks, which we identify as the mixed-phase (m-LLZO) (see Figure 2b). The split (211) peak in the tetragonal phase stems from the slight difference of the interplanar spacing between the (211) and (112) planes, which in the cubic structure becomes one single peak due to the identical lattice spacings (insets of Figure 2).

The LLZO pellets contain crystalline grains of about 1μm in size (see Figure S1). To get an insight into the microstructure of single grains, we perform synchrotron-based single grain diffraction, which leverages the high x-ray flux. We also exploit the high transverse coherence length of a third-generation synchrotron radiation source[25] to successfully perform BCDI on the LLZO grains embedded in the sintered pellet (see Figure 1b for a schematic experimental setup). The randomly oriented LLZO grains in a ~100 μm thick pellet are illuminated by the incident coherent X-ray beam at a photon energy of 10 KeV. The diffraction pattern around the (211) Bragg peak is recorded by a 2D detector positioned 2 meters downstream from the sample at a 2θ angle of 13.7 degrees around q = 1.2 Å$^{-1}$ (q=4π sin(θ)/λ, where λ is the wavelength). The random orientation of grains inside the pellet and the high angular sensitivity of Bragg diffraction allow us to isolate the coherent X-ray diffraction pattern from a single grain. We acquire the 3D reciprocal space map by rotating the sample stage in steps by about 1 degree (Figure S2). We first analyze the single particle diffraction patterns directly and then reconstruct the real-space 3D displacement field of the individual grains through an iterative phase retrieval algorithm.[26] We use the 3D images of the interior structure of LLZO to study crystal defects, particularly dislocations and strain gradients.[21–23,27]

While the powder XRD patterns in Figure 2 yield statistically averaged information, the single-grain diffraction provides a perspective on the microstructure of LLZO single grains. By measuring single-grain diffraction patterns of a dozen grains in LLZO pellets of each structural phase, we are able to conduct a statistical analysis on the microstructure of the individual LLZO grains (see Figure 3). In the absence of Al doping at x = 0, the LLZO grains present two distinct types of diffraction data: a split peak similar to the averaged powder XRD or a single peak (Figure 3a, see also Figure S3). Most grains in the tetragonal structure show only one diffraction peak (blue or yellow in Figure 3a), where the Bragg angle can be associated with either the (211) or the (112) peaks in the powder XRD (Figure 2a). The single peaks show that each of these individual grains consists of a single domain, oriented differently in different grains. The average over multiple grains yields the double peak structure observed in powder XRD. The second type of single-grain diffraction peaks, nevertheless, presents a split peak between q = 1.21 Å$^{-1}$ and 1.23 Å$^{-1}$, similar to the powder XRD but at the single grain level. The presence of a double peak indicates that a grain consists of two domains, one with (211) and the other with (112) planes aligned with the scattering vector and separated by a twin boundary inside single t-LLZO grains (see inset of Figure 3a).

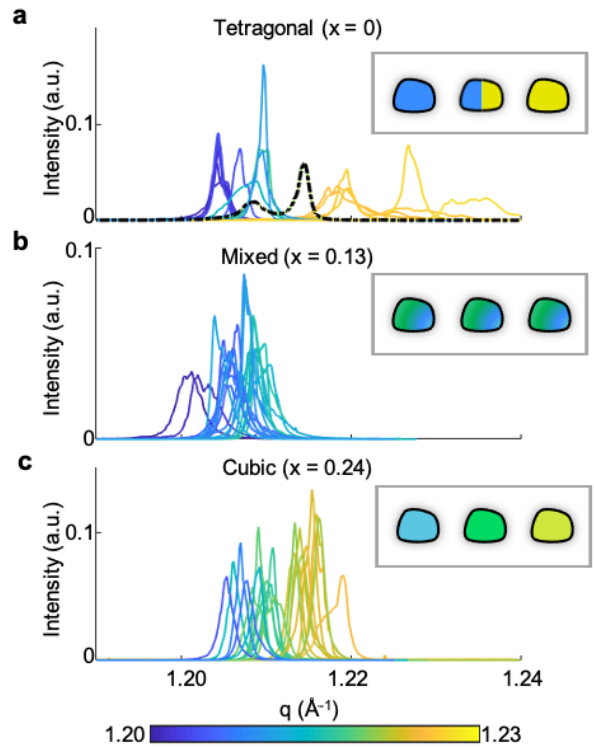

**Figure 3**. The x-ray diffraction intensity for individual LLZO grains in the (**a**) tetragonal, (**b**) mixed, and (**c**) cubic structure as a function of momentum transfer q. The colors of the peak indicate the peak position in q, and every peak-curve has one fixed color. The intensity is normalized by the integrated intensity of the peak (note the maximum intensity is higher in the cubic phase due to the shaper diffraction peaks in (**c**). In (**a**), two types of tetragonal peaks are visible: split peaks (one example is shown in dashed black for clarity) and single peaks at either low q (blue) or high q (yellow). The insets illustrate the possible constitution of crystal domains for each type of grain. Tetragonal grains can either be in the two uniform

crystal orientations, (211) as blue and (112) as yellow rotated by 90 degrees from the other or contain two coexisting domains. The mixed-phase grains all display a similar lattice constant, but the peak broadening indicates a strain gradient. The cubic structure has a uniform domain, but the lattice constant varies between different grains.

To investigate the structural morphology of LLZO due to the introduction of Al dopants, we analyze the Bragg peak position and shape. Following Bragg's law, the peak position in the reciprocal space is inversely proportional to the interplanar spacing, and a peak broadens with an increasing strain gradient in the grain.[29,30] Although both the c-LLZO and m-LLZO grains present only one peak in Figure 3b & 3c, the peaks of m-LLZO all center around $q = 1.21$ Å$^{-1}$ but are 50% wider. On the contrary, the sharper peaks in c-LLZO have a distribution with a 75% higher variation than the mixed structure (See Figure S4). The x-ray data allows the following interpretation. When the Al content is small, the limited substitution of Li sites with Al ions results in a large strain gradient, possibly due to the development of inhomogeneity of phases within the m-LLZO grains (see inset of Figure 3b). As the Al doping increases, a full phase transformation is activated, and large strain-free crystalline domains of stable c-LLZO form. Nevertheless, our data show that the lattice spacing in the grains varies slightly, likely due to the difference in the chemical composition of Al (see inset of Figure 3c).

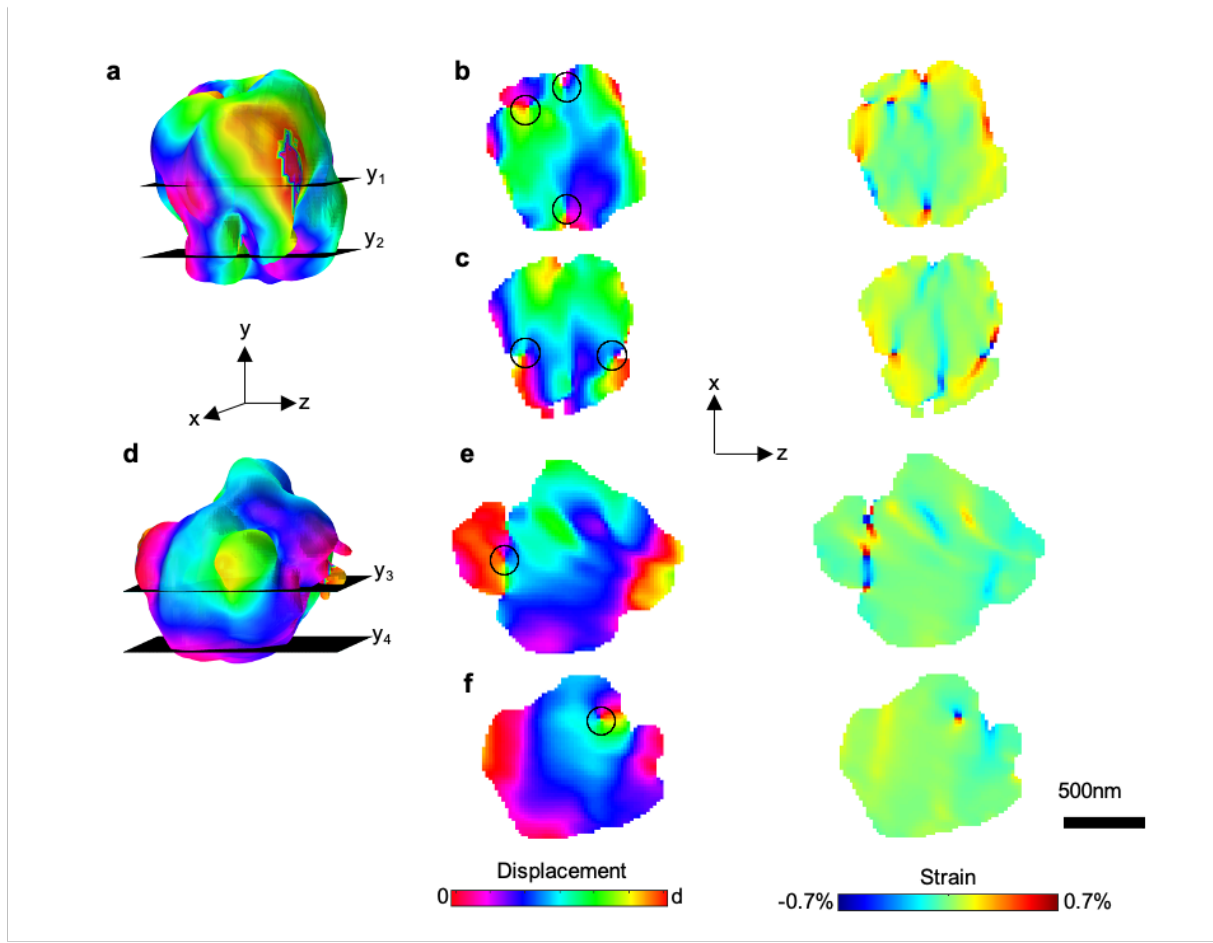

**Figure 4**. The reconstructed 3D displacement field of the LLZO grains with the (**a**) mixed and (**d**) cubic structure. The xz cross-section of the displacement (left) and the calculated strain (right) maps of the mixed structure at (**b**) $y_1$ and (**c**) $y_2$, and of the cubic structure at (**e**) $y_3$ and (**f**) $y_4$. Dislocations are visible as singularities in the displacement field. In the strain maps, the dislocation lines appear as regions of the compressive strain (blue) connecting with the tensile strain (red). The scattering vector (normal to the (211) lattice planes) points along the z-axis, and the scale bar is 500 nm.

To further investigate the type of extended defects within LLZO grains, the diffraction data (see Figure S5) is inverted through a rigorous phase retrieval algorithm that results in a three-dimensional complex-valued function.[26] The final result is an average of ten single reconstructions (see Figure S6). The complex function's amplitude is the density of the scattering planes (reflecting the shape of the grain), and the phase is the displacement field from the ideal lattice measured along the scattering vector q.[31] The strain is then calculated as the spatial derivative of the displacement field, which is connected with stress originating from chemical, mechanical, and other forces on the local environment of the crystal.[25] The reconstructed 3D shape and displacement field for the mixed and cubic structure of LLZO grains is shown in Figure 4a & 4d, where the cross-sections of the 3D phase are shown in the left of Fig 4b, c, e, and f. The 2D maps display singularities in the displacement field (circled in Fig 4 b, c, e, and f) associated with dislocations.[22] Interestingly, the reconstruction on single tetragonal grain does not show such features (see Figure S7).

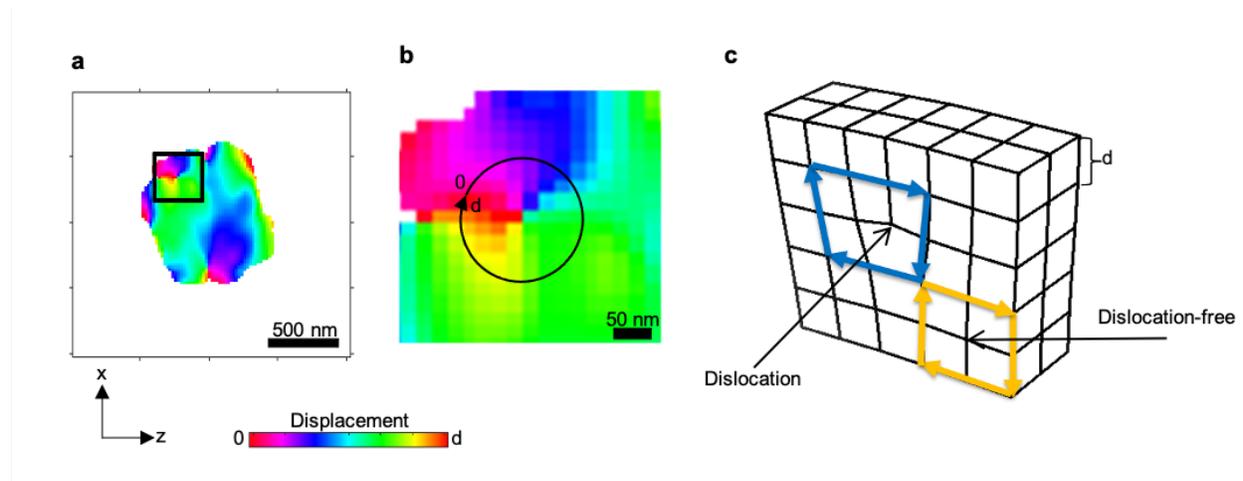

**Figure 5**. (**a**) The 2D slice of the 3D reconstructed displacement field. (**b**) The zoom of the black square in (**a**). At the singularity (center of the loop indicated by the black circle), the displacement field is discontinuous: it changes by one lattice spacing, d, when tracked along the loop. (**c**) A schematic of an edge dislocation and the schematic of a Burgers circuit. The displacement along the loop in a dislocation free crystal (yellow) is continuous, while the loop around a dislocation (blue) results in an extra spacing along the Burgers vector.

To evaluate the defect's type, we zoom in on the singularity in Figure 4b (see Figure 5). Along an arbitrary loop around the singularity, the displacement field changes from 0 to d (Figure 5b), which equals to one lattice spacing between, d, the (211) planes. The equivalence of the loop with the Burgers establishes the relationship between the singularity and the presence of dislocations inside the grains.[21–23,27] At a dislocation site, the Burgers circuit is one extra lattice spacing, d, longer than the Burgers circuit around a perfect crystal (Figure 5c). Since the experimental BCDI geometry is only sensitive the displacement along the scattering vector direction [211] and the displacement at the singularity is exactly one lattice spacing, we conclude that the Burgers vector is likely along the [211] direction. The dislocation lines directly visible in the 3D displacement field appear perpendicular to the Burgers vector (see Supplementary movies); thus, we identify all observed dislocations to be edge dislocations.[32] The dislocation lines are also visible in the strain maps (right of Fig 4b, 4c, 4e, and 4f), where the tensile strain connects with the compressive strain (in blue and red).

The reconstructed displacement maps and the calculated strain fields in Figure 4 show more dislocations in the mixed than in the cubic structure (more 2D slices shown in Figure S8, also see movies in Supplemental Information). In addition, the particle in the cubic structure is mostly strain-free except around the dislocation lines, while the mixed-phase grain shows extended regions of tensile strain (yellow region in Figure S9) adjacent to the dislocations. The lattice spacing in the extended regions is about 0.2% larger than in the rest of the grain. The difference in the lattice constant of the mixed structure is much smaller than the 2% difference of the lattice spacing between the (112) and (211) planes of the tetragonal structure. The difference in m-LLZO is more likely due to the reduction of tetragonality from Al doping and the inhomogeneous Al concentration. The observation of these strained nanodomains revealed by the 3D imaging agrees with our previous conclusion of a large strain gradient in m-LLZO from the single-grain diffraction, which can lead to extensive phase boundary.

To substantiate the difference in the microstructure between the cubic and mixed structures, we calculated the partial strain energy along the [211] direction $E_{211} = \frac{1}{2}Y \int [\varepsilon_{211}(v) - \text{mean}(\varepsilon_{211}(v))]^2 dv$, where Y is the Young's modulus of the material and $\varepsilon_{211}(v)$ is the 3D measured strain, and the sum is taken through the volume of the particle v.[23] Using 150 GPa as the modulus for both structures,[33] the reconstructed particle in the mixed structural phase has a strain energy of 8.5 nJ/μm³, twice higher than the $E_{211}$ of the particle in the cubic phase of 4.0 nJ/μm³. The strain energy corroborates our previous discussion based on the phase and strain maps: namely, the mixed structural phase particle shows more singularities and domain structures that result in high strains.

The microstructure of c-LLZO and m-LLZO characterized by the BCDI inherently influences materials' mechanical properties and likely has implications on its electrochemical performance as a solid-state electrolyte. Studies have shown that the microstructural inhomogeneity - dictated by the structural defects such as the dislocation networks - can affect the nucleation tendency of Li accumulation.[34–36] Thus, we identify the dislocation sites that we reveal with BCDI as areas prone to the formation of Li dendrites, a significant challenge of using Li metal batteries.[37] Furthermore, fracture toughness, a quantitative property that reflects the resistance of cycling-induced fracture, is heavily dependent on materials' microstructural defects such as grain size, impurities, and pre-existing cracks.[1,38] Studies have modeled that at a high fracture toughness, the creation and propagation of cracks can be avoided, which results in stable battery performance.[39]

The inhomogeneous residual strain field that arises from the dislocations in the mixed and cubic structures has implications on the structural integrity of the grains. If dislocations cannot propagate at low temperatures due to the high Peierls stress in the brittle ceramic LLZO, the mixed structure could fail through inter- or intra-granular fracture, and the high and inhomogeneous residual stresses could accelerate this process.[40] The residual stresses at the unloaded state minimize the range of allowable applied stresses that the grain can withstand without exceeding the ultimate strength of the material. In the cubic structure, the severity of the strain inhomogeneity is not as extensive since it has fewer dislocations, which could translate positively towards its structural integrity in extreme loading conditions.

The microstructure also directly relates to the ionic transport in Al-doped LLZO. Extended crystalline defects such as grain boundaries have been reported to suppress ionic diffusion inside LLZO.[28,41] Therefore, the twin boundaries we report inside t-LLZO grains could also limit the ionic transport and reduce the overall conductivity. Computational studies show that tensile strain, which we observe in the reconstructed m-LLZO, would reduce the ionic conductivity by a factor of 2.[42] However, other studies have also predicted that the tension of lattice parameters can instead increase the ionic conductivity of Al-LLZO due to the expansion of the triangle diffusion bottleneck.[18,43] Despite the uncertainty on the exact effect of strain on Al-LLZO, we anticipate that the inter- and intra-particle strain heterogeneity that we observe in the mixed and cubic structure with BCDI has a sizable impact on its ionic conductivity.

In summary, we used single grain diffraction and BCDI to study the structural heterogeneity and extended crystal defects of LLZO at various degrees of Al doping. We observed twin domains inside undoped single tetragonal-LLZO grains. We also found that low Al doping results in a large strain gradient in mixed-LLZO, and as the doping increases, cubic-LLZO grains stabilize with a reduced strain gradient but slightly different average lattice constants likely as a result of varying Al concentration. The reconstructed displacement field of both the mixed-LLZO and cubic-LLZO single grains reveals edge dislocations. In proximity to dislocations, the mixed-LLZO structure also exhibits extended tensile strain regions that indicate subdomains of another structural phase separated by an extensive grain boundary, including dislocations. The presence of crystal defects reported here shows the critical role that Al doping plays in modifying the microstructure of LLZO. In the future, a combination of operando spectroscopy and imaging techniques is required to better quantify the connection between structural defects to ionic transport and cycling stability of solid-state electrolytes.

## Author Contributions

A.S. and F.L. conceived of the idea. L. M. and F.L. prepared the samples and performed the X-ray powder diffraction. O.G., D.W., R.B. and W.C. performed the synchrotron-based measurements. Y.S analyzed the data and wrote the paper. All authors participated in the interpretation of the data and revised the manuscript.

## Acknowledgments

We thank Dr. L. Archer for discussions. The work at Cornell was supported by the NSF CAREER Award Grant 1944907. The work at Virginia Tech was supported by the National Science Foundation (Grant no. DMR 1832613). This research used resources of the Advanced Photon Source, a U.S. Department of Energy (DOE) Office of Science User Facility, operated for the DOE Office of Science by Argonne National Laboratory under Contract No. DE-AC02-06CH11357.

# Supplemental Information for

# X-ray nanoimaging of crystal defects in single grains of solid-state electrolyte Al$_x$Li$_{7-3x}$La$_3$Zr$_2$O$_{12}$

*Yifei Sun[1], Oleg Gorobstov[1], Linqin Mu[2], Daniel Weinstock[1], Ryan Bouck[1], Wonsuk Cha[3], Nikolaos Bouklas[4], Feng Lin[2,\*], Andrej Singer[1,\*]*

[1]*Department of Materials Science and Engineering, Cornell University, Ithaca, New York 14850, United States*
[2]*Department of Chemistry, Virginia Tech, Blacksburg, Virginia 24061, United States*
[3]*Advanced Photon Source, Argonne National Laboratory, Argonne, Illinois 60439, United States*
[4]*Sibley School of Mechanical and Aerospace Engineering, Cornell University, Ithaca, New York 14850, United States*

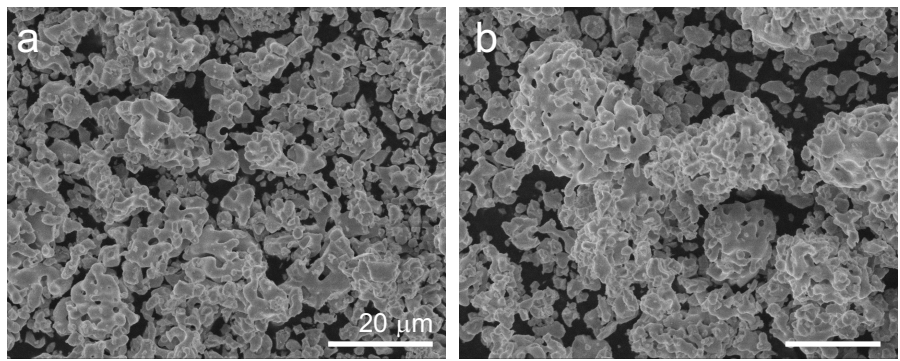

**Figure S1**. SEM images of the (**a**) tetragonal-Li$_7$La$_3$Zr$_2$O$_{12}$ and (**b**) cubic-Al$_{0.24}$Li$_{6.28}$La$_3$Zr$_2$O$_{12}$.

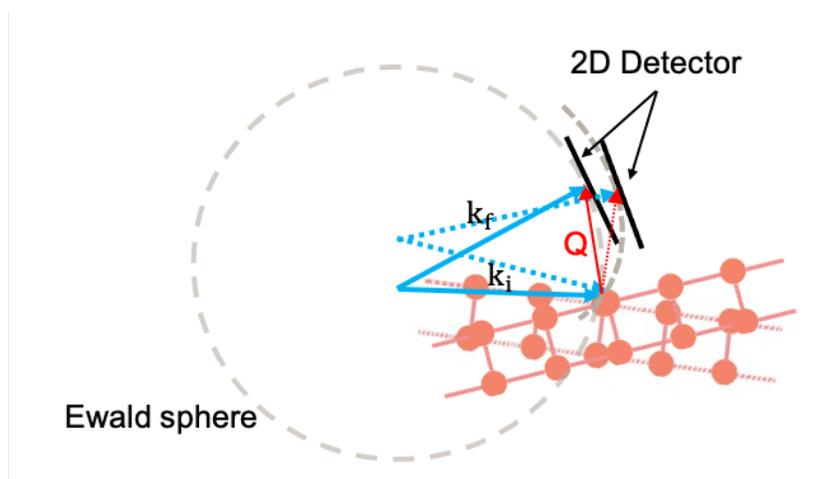

**Figure S2**. The momentum transfer q is given by the difference between the outgoing and incoming X-ray and the Laue condition requires that q has to equal to the reciprocal space vector. Since in elastic scattering, q is constrained to the Ewald sphere, we can only observe diffraction intensity when the Ewald sphere crosses with the reciprocal space vector. To acquire a 3D diffraction data, we need to rotate the sample so that the detector can record all of the 2D sections of the data.

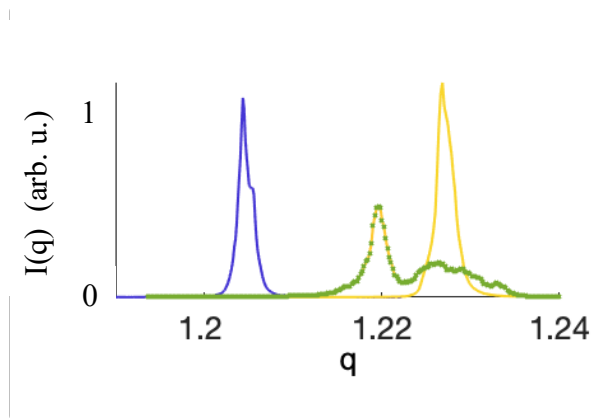

**Figure S3**. Two types of tetragonal peaks from single particle diffraction. Single peaks at either low q or high q (blue and yellow), and a split peak (dashed green).

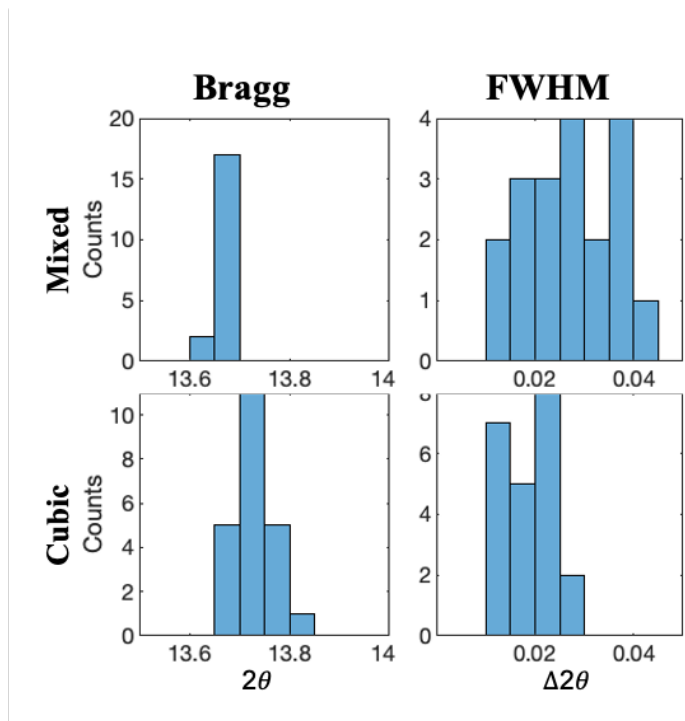

**Figure S4**. Histograms of Bragg angles (left) and peak widths (right) from the coherent single particle diffraction of all LLZO particles in the mixed and cubic structure. The standard deviation of the Bragg angle for the mixed structure is 0.043° and the cubic structure is 0.076°. The average width of the mixed structure is 0.027° and of the cubic structure is 0.018°.

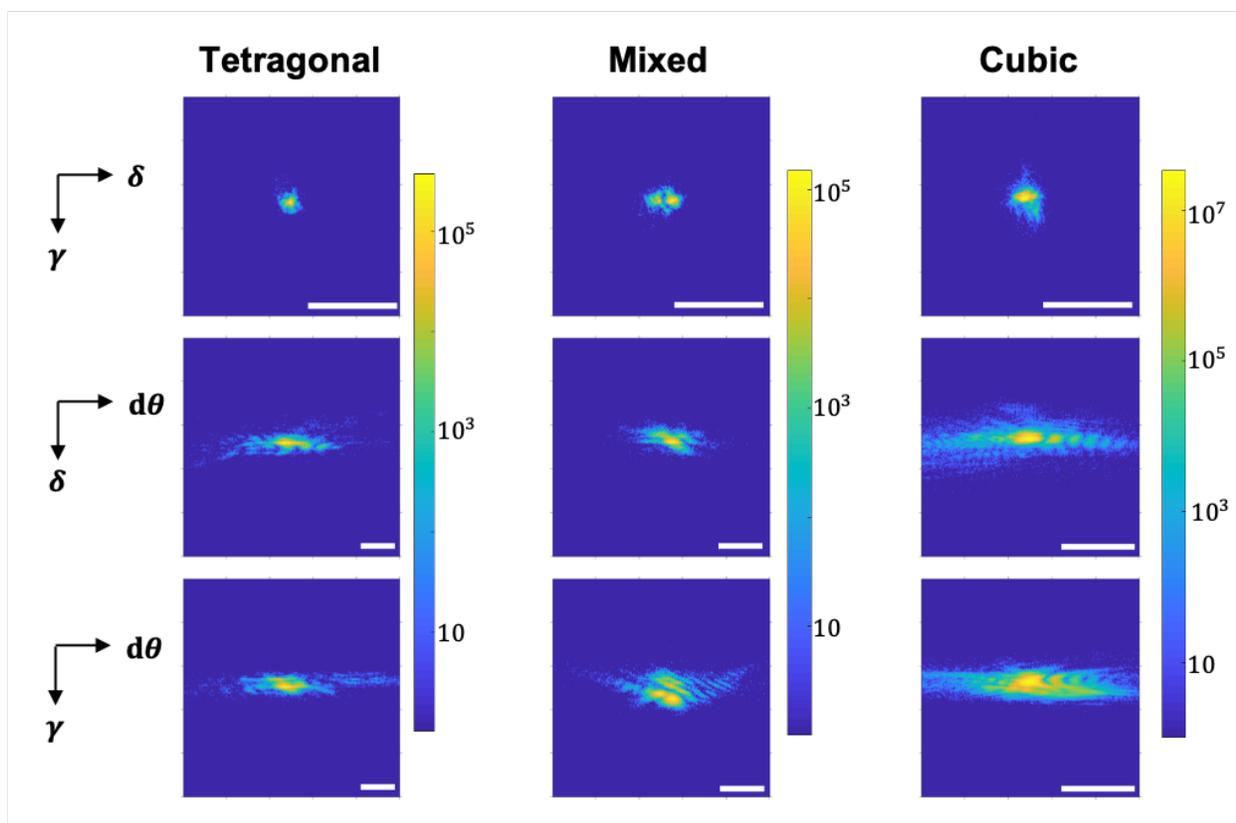

**Figure S5**. The orthogonal 2D slices of the 3D diffraction pattern at the maximum intensity for the tetragonal, mixed and cubic structure. γ and δ are the rotation axes of the detector. Both the mixed and cubic structure show asymmetric clear interference that allow reliable phase retrieval (see Fig. S6). The cubic structure has the highest intensity, two orders of magnitude higher than the tetragonal and mixed. Although the maximum intensity of the mixed and tetragonal structure is around the same magnitude, the integrated intensity of the mixed structure is 60% higher than the tetragonal structure. The scans are cropped to 128*128 pixels and binned to the one-third of original rocking scans to reduce the phase retrieval running time. The iterative phase retrieval algorithm adopts the guided approach with 40 populations in 8 generations of alternating ER and HIO methods for 610 iterations.[25,44] For each grain, 10 individual runs of reconstruction are performed and averaged, where each starts with a different random phase. Here we use the phase retrieval code developed by Jesse Clark and others (see for example, Ref. 26 in the main text) and widely used for reconstructing experimental data recorded at 34ID-C beamline of the APS.

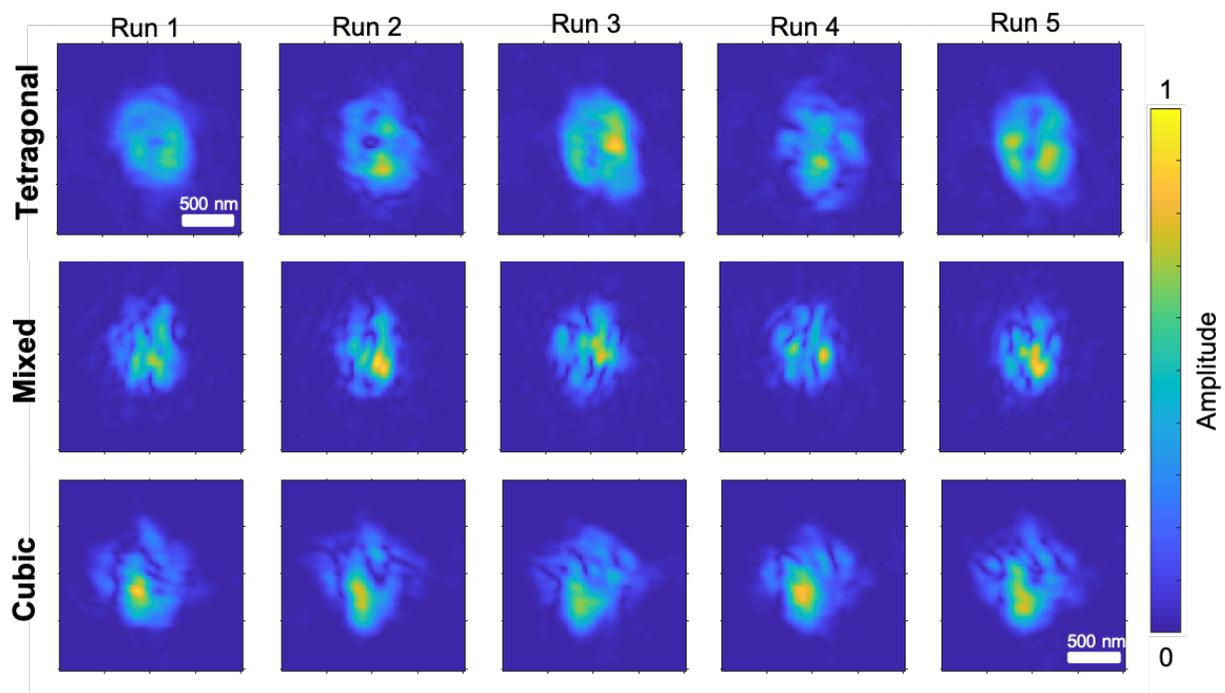

**Figure S6**. The reconstructed amplitude of the tetragonal, mixed and cubic structure of single LLZO grain. Because the iterative phase retrieval is model-independent and always converges to a solution, no rigorous test for the validity of this solution exists. The research of the last decade has shown, that a solution is correct if multiple phase retrieval runs, all started from different initial random phases, converge to the same solution.[22] The validity of the reconstructions is based on persisting features across different runs. The reconstructions on the cubic structure are thus trustworthy due to the consistency of the shapes and the location of defects. The validity of the tetragonal and mixed structure can be confirmed in the phase maps in **Figure S7**.

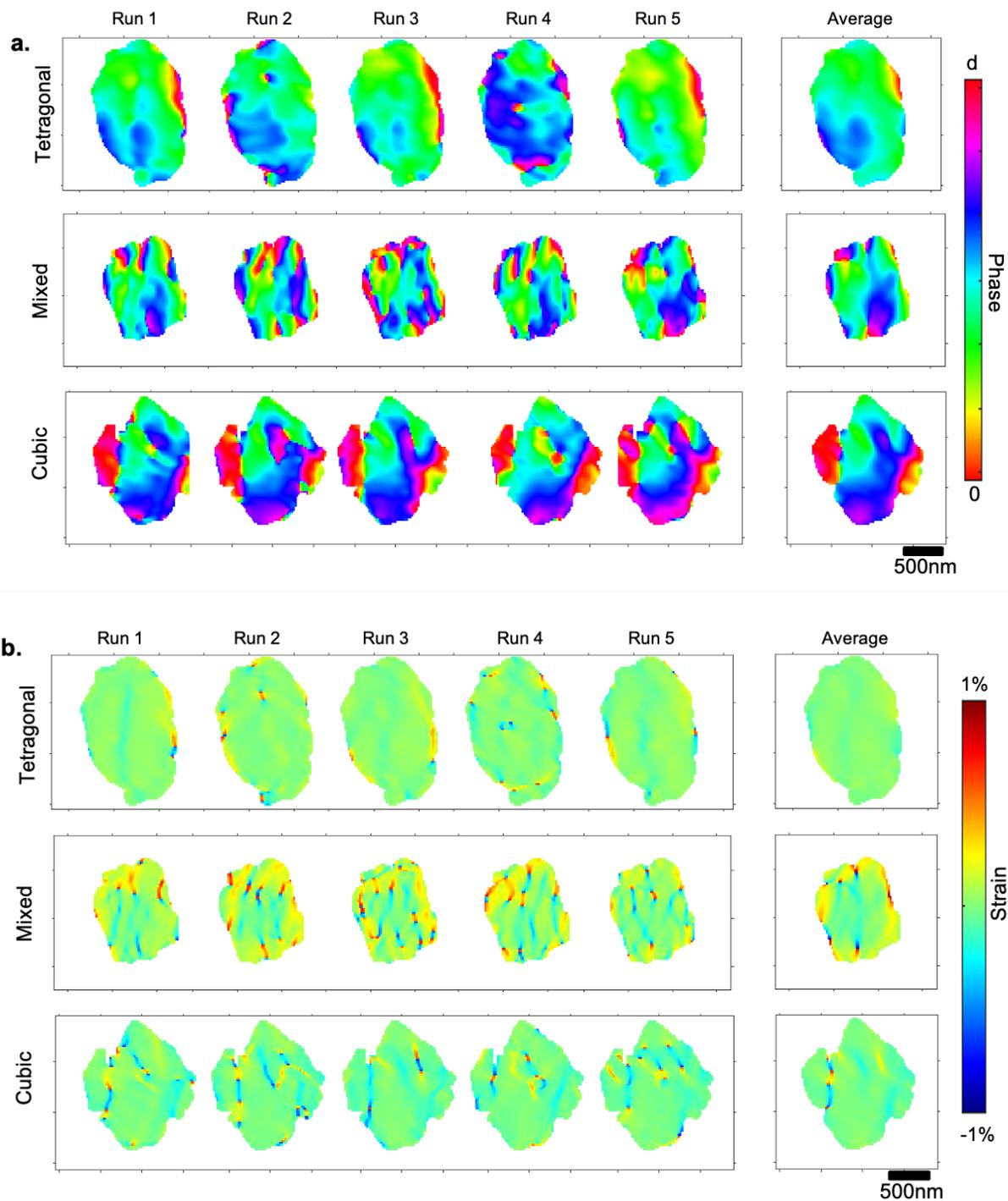

**Figure S7**. The (**a**) reconstructed displacement field and (**b**) calculated strain field of the LLZO single grain in the tetragonal, mixed, and cubic structure. On the displacement maps in (**a**), all three structures show consistent results from individual reconstructions and the averaged reconstructions also show similar features. However, on the strain maps in (**b**), although the defects with higher strain values persist across separate individual reconstructions and in the averaged results for the mixed and cubic structure, the tetragonal shows no defect and is entirely strain-free in the averaged result.

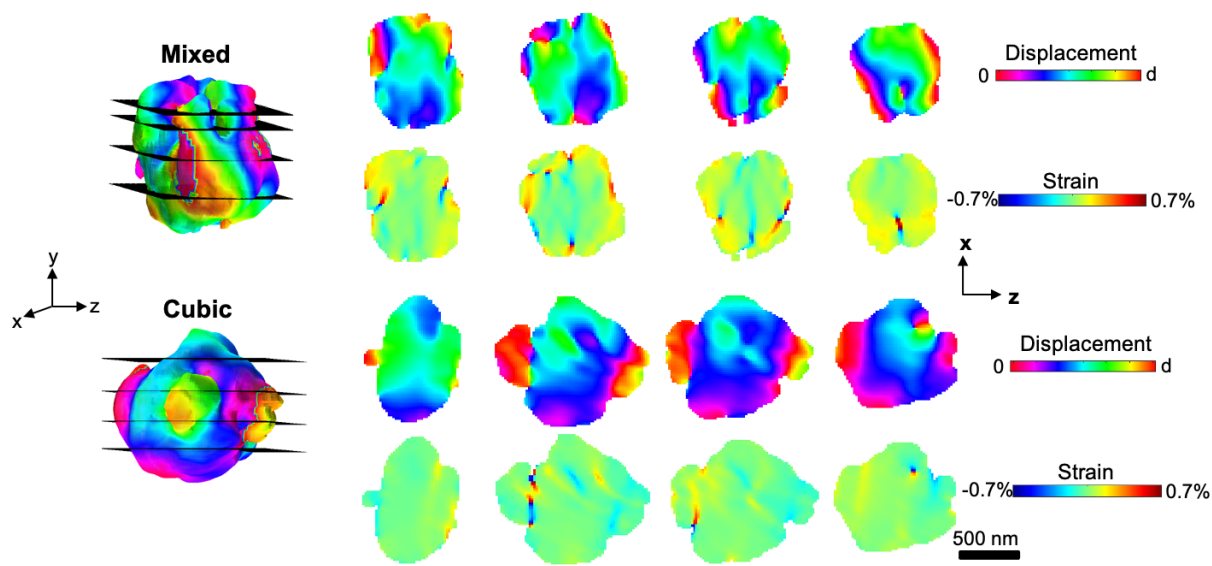

**Figure S8.** Reconstructed 3D displacement field and the cross sections of the displacement field (top) and calculated strain field (bottom) for the mixed and cubic structure. Z-axis points along the scattering momentum direction. The mixed structure shows more defects around the edge of the particle, while the cubic shows less defects that penetrate deeper within the particle.

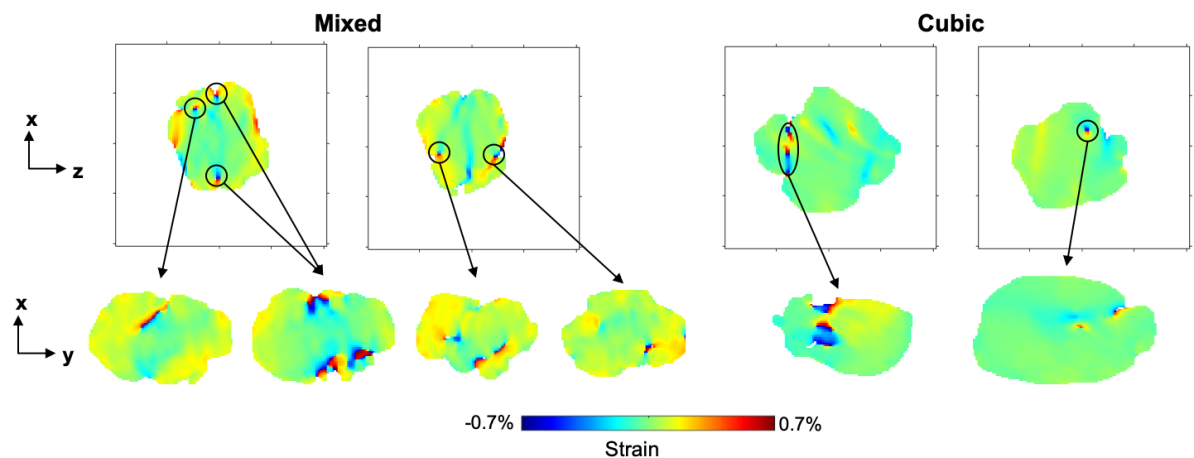

**Figure S9.** Additional side views of the cross sections of the strain field in Figure 4. The x-y plane further illustrates the sub-domains of different structural phase (positive strain in yellow) that nucleate near the dislocations in the mixed structure, whereas the cubic structure is almost strain-free (green) except at the dislocations.